\documentclass[twocolumn,showpacs,preprintnumbers,amsmath,amssymb]{revtex4}

\newcounter{ctr}

\usepackage{graphicx}
\usepackage{epsfig}
\usepackage{dcolumn}
\usepackage{bm}

\begin{document}

\title{Conditions for Chemotactic Aggregation}

\author{Masayo Inoue$^1$ and Kunihiko Kaneko$^{1,2}$}
\affiliation{
$^1$Department of Basic Science, Graduate School of Arts and Sciences, University of Tokyo, \\
3-8-1 Komaba, Meguro-ku, Tokyo 153-8902, Japan \\
$^2$ ERATO Complex Systems Biology Project, JST,  \\
3-8-1 Komaba, Meguro-ku, Tokyo 153-8902, Japan \\ }
\date{\today}

\begin{abstract}
Micro-organisms aggregate through chemotaxis against a concentration gradient of signals secreted by themselves. We have numerically studied a model consisting of elements with intracellular dynamics, random walks with a state-dependent turnover rate, and secretion of attractant. Three phases with and without aggregation, as well as partial aggregation, were obtained as to the diffusion and decomposition rates of the attractant, and conditions for cellular aggregation were analyzed. The size of aggregated clusters was shown to be independent of cell density, as is consistent with experiment.
\end{abstract}


\maketitle

Chemotaxis is a ubiquitous phenomenon in microorganisms, and has attracted much attention both from the experimental and theoretical sides\cite{Oosawa_JTB77, BergBrown_nature72, Koshland_PNAS72, Leibler_CRAS01, Clark_PNAS05}. The external concentration of signal molecules is interpreted by an intracellular signal transduction network, which changes the motility of the cell, so that it moves toward a region with a higher concentration of the attractive signal molecule\cite{Koshland_PNAS72}. The signal pathways governing chemotaxis have been revealed experimentally\cite{Sourjik_TM04}. In bacteria, $\textit{Escherichia coli}$, the turnover rate for the random walk is modulated by the signal concentration toward the directed motion on average\cite{Koshland_science73, Koshland_PNAS75}. This is also true for several micro-organisms together. In fact, from experiments on $\textit{Paramecium}$, Oosawa and Nakaoka proposed a condition for chemotaxis, which states that the time scale of tumbling must be smaller than that of adaptation and greater than that of sensing\cite{Oosawa_JTB77}. By using a simplified model for internal signal transduction and random turnover, we have recently confirmed that the condition is valid for a variety of environments and for both short- and long-term behavior by suitable renormalization of the parameters for the timescale\cite{MIKK_PRE06}.

Just as the chemotaxis of a single microorganism is of interest, the collective chemotaxis of microorganisms interacting with each other is also of interest\cite{Benjacob_PRL95, Matsushita_PhysicaD05}. For example, $\textit{E. coli}$ aggregate to form a cluster by using chemotaxis\cite{Mittal_pnas03} or sometimes generate complex patterns\cite{BudreneBerg_nature91}. The aggregation is spontaneous, the result of chemotaxis toward a chemical that is secreted by the bacteria themselves. Recently, Mittal et al.\cite{Mittal_pnas03} studied this chemotactic aggregation and found that the size of the bacterial cluster is independent of the number of bacteria therein. Some analysis was performed by imposing the localized signal pattern in advance\cite{Mittal_pnas03}.
However, such a concentration pattern is generated by the aggregating cells themselves, and thus it is essential to obtain a self-consistent condition between the bacterial distribution and the signal field to allow for chemotactic aggregation. In the present Letter, we will study a simple model of elements that show chemotaxis and secrete signal molecules, in order to obtain the conditions for chemotactic aggregation. Dependence of the cluster size on the bacterial number will also be examined.

Our model consists of cells with internal chemical reactions showing response to and adaptation against the signal molecule\cite{AsakuraHonda_JMB84}; the turnover rate of the random walk of cells depends on the internal chemical state, while the speed of motion is fixed at $v_{speed}$ for simplicity. Signal molecules are secreted from the cells into the medium, become diffused, and are decomposed. The intracellular process for chemical concentration variables $c_u$ and $c_v$ is based on\cite{ErbanOthmer_SIAM04, MIKK_PRE06}. These chemicals respond to the external signal concentration $S$, and the intracellular adaptive dynamics is represented by 
\begin{eqnarray}
\frac{dc_u}{dt}=\frac{S- (c_u + c_v)}{\tau_s } , \frac{dc_v}{dt}=\frac{S- c_v}{\tau_a }.
 \label{eq_othmer}
\end{eqnarray}

Following the increase (decrease) in the signal concentration $S$, $c_u$ increases (decreases) from its steady state value ($c_u^* =0$\cite{com1}), but after some time span 
it returns to the original value $c_u^*$. The timescale for the response is given by $\tau_s$, while that for the relaxation to the original value (i.e., adaptation) is given by $\tau_a$. 

Following the experimental result\cite{Koshland_science73, Koshland_PNAS75}, we set the tumbling rate to become smaller when $c_u > c_u^*$ and larger when $c_u < c_u^*$.
With the average tumbling time-interval $\tau^*$ and the speed $v_{speed}$, we set the tumbling probability (per unit time) as
\begin{eqnarray}
 P_{tmb}(c_u) = \frac{1.0 - 0.5 \times \tanh(\kappa_{\Delta} (c_u-c_u^*))}{\tau^*}. 
  \label{eq_tumble}
\end{eqnarray}
The tumbling frequency decreases (increases) as $S$ increases (decreases).
Unless otherwise mentioned we choose $\kappa_{\Delta}=1000.0$, so that the $P_{tmb}(c_u)$ exhibits a threshold behavior.
Although the tumbling occurs randomly, this simple model can show chemotaxis: i.e., cells move toward an attractant-rich area under the condition $\tau_s<\tau*<\tau_a$, which we term the Oosawa condition\cite{MIKK_PRE06}. (The response time $\tau_s$ and adaptation time $\tau_a$ need to be properly rescaled depending on the profile of signal concentration\cite{MIKK_PRE06}).

Now we consider the process of secretion of signal molecules $S$ by the cells, to consider the spontaneous chemotactic aggregation. The chemical is assumed to be secreted continually with a constant rate $\sigma$ from each cell, diffuses through the space with the diffusion coefficient $D_s$, and is decomposed at the rate $\nu$. Thus, the time evolution of the signal concentration is given by
\begin{equation}
 \frac{\partial S(x,t)}{\partial t} = \sigma \sum_{i}^{N_{cell}} \delta (x -x_{i}^{cell} ) + D_{s} \frac{\partial^2 S}{\partial x^2} -\nu S .
   \label {eq_S_dynamics}
\end{equation}

When cells are distributed homogeneously in space, the signal concentration approaches a homogeneous steady state $S^{*}=\sigma \rho /\nu $ with $\rho $ density of cells $N_{cell}/L$, $N_{cell}$ as the number of cells and $L$ as the system size\cite{remark}.

In our model, the concentration pattern of the signal chemical changes over time, influenced by the configuration of cells. On the other hand, cells move according to the signal pattern.
Cells regulate the signal pattern, which controls the cells' motion. Chemotactic aggregation is possible, when a stationary self-consistent solution between cells' motion and the time evolution of the signal pattern is realized.

If the signal concentration and cell density change smoothly in space, and each cell's adaptation dynamics is averaged out to consider only the distribution of cells, it would be possible to make a coarse-grained description at suitable temporal and spatial scales. Indeed, Erban and Othmer derived a partial differential equation considering a continuum approximation for chemotactic particles under suitable conditions\cite{ErbanOthmer_SIAM04}. The derived equation agrees with the so-called Keller-Segel model\cite{Keller_70}, originally introduced for the study of chemotactic aggregation of amoeba\cite{Nanjundiah_73}. By denoting the cell density in space
as $N(x,t)$, the derived equation is written as
\begin{eqnarray}
\frac{\partial }{\partial t} N(x,t)= - \frac{\partial }{\partial x} \Bigl[ \chi \frac{\partial S(x,t)}{\partial x}  
 - D_n \frac{\partial }{\partial x} \Bigr] N(x,t) ,\nonumber \\
\frac{\partial }{\partial t} S(x,t)= \sigma N(x,t) -\nu  S(x,t) + D_s \frac{\partial ^2}{\partial x^2} S(x,t).
 \label{eq_keller}
\end{eqnarray}
\noindent
where $\chi$ represents the mobility of the cell against the signal gradient, and $D_n$ is the diffusion of cells due to their random walk. In this continuum limit, cells are assumed to show directed motion even at any slight gradient in the signal chemical. From a straightforward linear stability analysis, it is shown that the Keller-Segel model has a steady uniform solution under the condition of $\nu > \nu_c =\sigma \chi N^* / D_n $, where $N^*$ is given by the density of cells $\rho $. In the one-dimensional case, Childress and Percus obtained a stationary localized solution for $\nu < \nu_c$, which represents the aggregation\cite{Childress_81}.

Here we consider the cell model in one-dimensional space, without taking a continuum limit, and study the conditions for chemotactic aggregation, in particular dependence on $D_s$ and $\nu$. The parameters for intracellular dynamics, i.e., $\tau_s, \tau_a, \tau^*$, are fixed so that they satisfy the Oosawa condition: i.e., at the single cell level, chemotaxis is possible. Unless otherwise mentioned, the number of cells is 100. Although we present simulations of the one-dimensional case only, the preliminary results suggest that the basic properties, such as the conditions for aggregation and the cluster size, are invariant even for the two-dimensional case, as adopted experimentally. 

\begin{figure}
\begin{center}
\scalebox{0.4}{\includegraphics{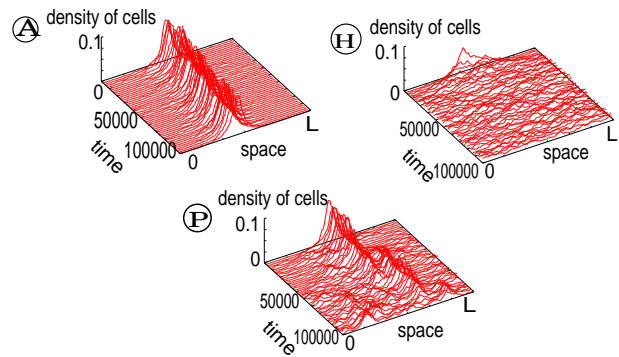}}
\caption{(Color online) Three characteristic behaviors of cells. The time evolution of the density distribution of cells is plotted. The distribution is computed by averaging over 1,000 time units. Parameters are commonly set as $\tau_s= 5.0, \tau_a=50.0, \tau^*=70.0, \sigma = 0.0005$, $N_{cell}=100$, and $L_=3000$, while $D_s$ and $\nu$ are chosen to be (\textsl{A}) $ D_s=300, \nu =0.002 $, (\textsl{P}) $ D_s=0.1, \nu =0.002$, and (\textsl{H})$ D_s=300, \nu = 0.1 $.}
    \label{pic_cluster_sozu}
\end{center}
\end{figure}

By fixing the parameter values of the intracellular process, we studied the temporal evolution of distribution by changing the parameter values $ D_{s} $ and $ \nu $, and found three distinct types of behavior, i.e. aggregation(A), homogeneous distribution(H), and partial aggregation(P), as shown in Fig.\ref{pic_cluster_sozu}.

At the aggregation phase, cells aggregate into a single cluster, which is localized in space and stable in time. This single cluster is formed irrespective of the initial distribution of cells. At the partial-aggregation phase, cells aggregate to form a cluster for some time span, but then this cluster collapses so that cells are broadly scattered until they aggregate again.  Intermittent aggregation and collapse is repeated. At the homogeneous phase, cells are distributed uniformly over the space. Tiny fluctuations in cell density are evident from time to time, but on the average the density is uniform in space.

\begin{figure}
\begin{center}
\scalebox{0.30}{\includegraphics{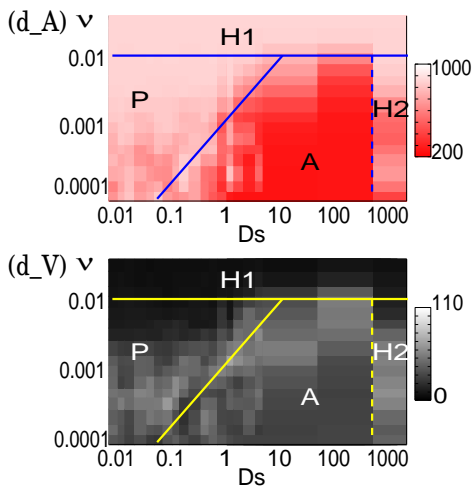}}
\caption{(Color online) Phase diagram with regards to the diffusion constant of signal chemical $ D_s$ (abscissa axis) and its decay rate $\nu $ (ordinate axis). Density plots of  $d_A$ (upper) and $d_V$ (lower) are shown. Four phases, H1, P, A, and H2, were obtained from these values. Parameters other than $D_s$ and $\nu$ are identical as adopted in Fig.1. The values $d_A$ and $d_V$ are computed from the average of 50,000 to 100,000 time steps, by starting from a homogeneous distribution.  }
    \label{pic_sozu_nu_ds}
\end{center}
\end{figure}

To characterize these behaviors, we computed the following two quantities: $d_A$, which characterizes the average spatial inhomogeneity of cells at a particular time, and $d_V$, which characterizes the temporal variation of cell aggregation. These are measured from the average cell-cell distance at each time
\begin{eqnarray}
   d(t) = \sqrt{ \frac{1}{N_{celll} (N_{cell}-1) } \sum_{i,j}^{N_{cell}} \| x_{cell}^i (t) - x_{cell}^j (t) \| ^2 }.
 \label{eq_dist}
\end{eqnarray}
\noindent
Then, $d_A$ is defined by the temporal average of $d(t)$ and $d_V$ by its temporal variance. The three phases are characterized by (\textsl{A}) $d_A << d_A^{uni}, d_V \sim 0$; (\textsl{P}) $d_A \sim d_A^{uni}, d_V>0 $; and (\textsl{H}) $d_A \sim d_A^{uni}, d_V \sim 0 $ where $d_A^{uni}= L/(2\sqrt{3})$, the value when cells are uniformly distributed. The parameter dependence of these quantities is plotted in Fig.\ref{pic_sozu_nu_ds}, from which the phase diagram is obtained. The diagram consists of four regions: i.e., aggregation(\textsl{A}), partial aggregation(\textsl{P}) and two regions of homogeneous phases(\textsl{H1},\textsl{H2}). Now, we discuss the transitions among these phases.

Recall that there are only two phases in the continuum limit, i.e., in the Keller-Segel model.
The boundary is given by $\nu = \nu_c$, beyond which the uniform solution of cell density and signal concentration is stable. This boundary line agrees with that separating the H1 phase and the other three phases in our model. In fact, in \textsl{H1}, the signal decay rate is too large to keep a sufficient signal amount for cells to detect. Since the aggregation cluster is always stable under $ \nu < \nu_c $ in the Keller-Segel model, the \textsl{P} and \textsl{H2} phases are a result of cell dynamics uncovered by the continuum limit.

The boundary between \textsl{A} and \textsl{P} is given by the straight line of $\nu \propto D_S$. Considering eq.(\ref{eq_S_dynamics}), the spatial scale for a signal molecule to diffuse within its lifetime ($\lambda _s $) is given by $\sqrt{D_s / \nu}$. Each cell has to respond to the signal change within this spatial scale. Now, we define the spatial scale for the cell's motility $\lambda _n $ as the average length a cell moves before it tumbles after it passes the central top of the signal field. This is estimated as follows. The cell's response against the change in signal concentration requires the time delay of $\tau _s$. Up to this time scale, the tumbling frequency does not change and cells seldom tumble. Since the tumbling probability is given by $1/ \tau^*$ per unit time, cells show diffusion going straight for the time span of $\tau*$, on the average. Hence, before the response to the signal the cells travel with the scale $\lambda _n \sim v_{\textit{speed}} (\tau_s + \sqrt{2} \tau^* ) $ on average\cite{eq_poisson}.

\begin{figure}
\begin{center}
\scalebox{0.27}{\includegraphics{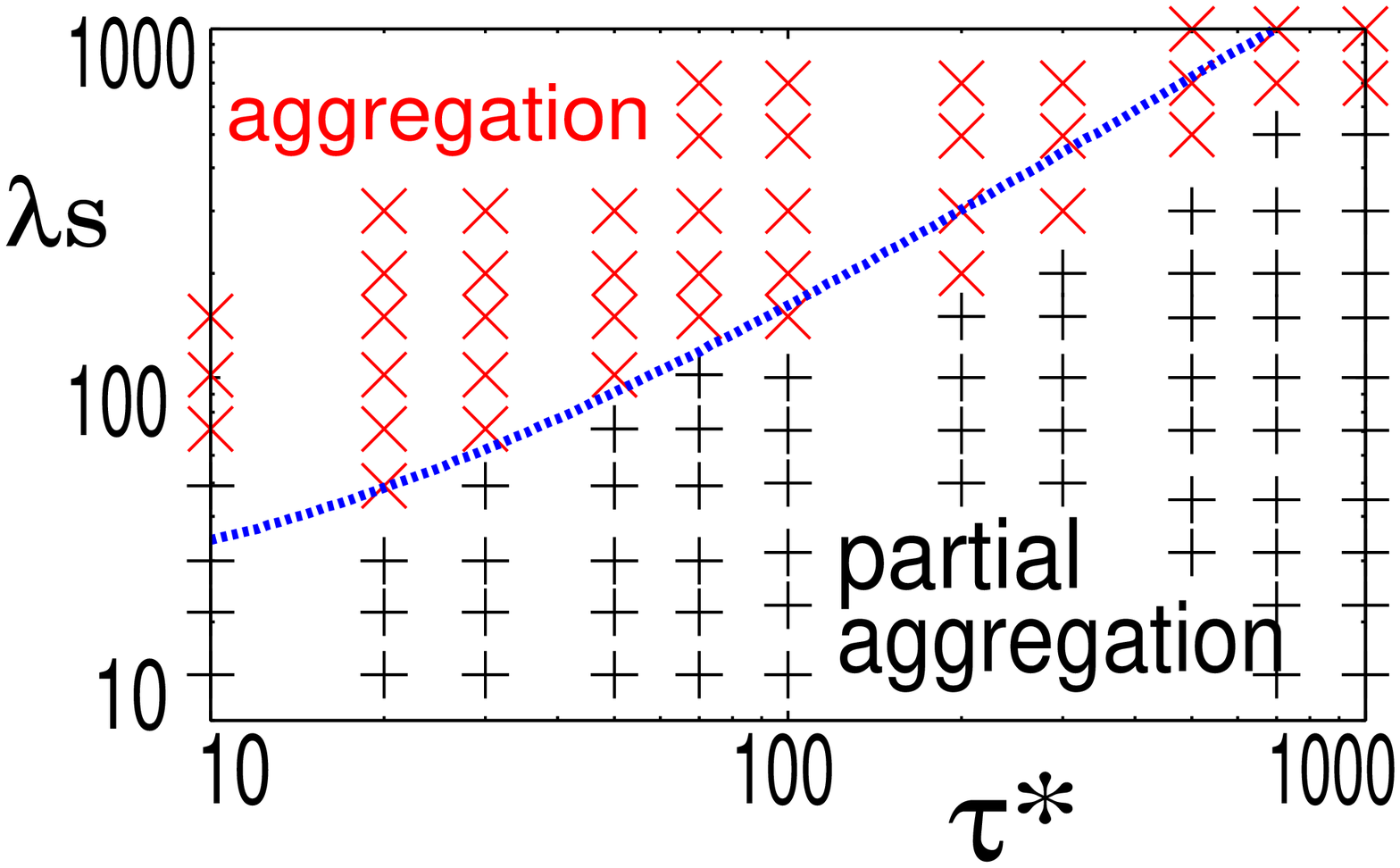}}
\caption{(Color online) Phase diagram of aggregation and partial aggregation phases, with regards to the cellular tumbling timescale $\tau^*$ (abscissa axis) and the parameter $ \lambda _s = \sqrt{D_s / \nu } $ (ordinate axis). The symbol $\times$ represents the aggregated phase ($d_A < d_A^{uni}$) and $+$ the partial aggregation phase($d_A > d_A^{uni}$). We set $\sigma=0.5$, while other parameter values are identical with those adopted in Fig.\ref{pic_sozu_nu_ds}.The dashed line represents $\lambda_n$ (see the text).}
    \label{pic_sozu_lums_tau}
\end{center}
\end{figure}

For a cell to respond to the change, the spatial scale of the signal change should be larger than the average length of the cell motion before response. Thus, the condition $\lambda _s > \lambda _n $ is imposed. This gives the boundary between the A and P phases in Fig.2, while we have explicitly confirmed the relationship between $\lambda_s$ and $\tau*$, as shown in Fig.\ref{pic_sozu_lums_tau}.  

When the aggregation condition in the continuum model ($\nu <\nu_c$) is satisfied but $\lambda _s < \lambda _n $, the signal field once formed cannot trap cells within, and they wander out so that the original cluster is destabilized. This leads to intermittent formation
and collapse of clusters. This is nothing but the behavior in the partial aggregation phase.
Note that in the continuous Keller-Segel model, there is always a drift in the cell motion towards a region with higher signal concentration, and the P-phase does not exist. By considering each cell as a discrete element with response by internal dynamics, the instability of the aggregated cluster under $\lambda _s < \lambda _n $ is introduced.

The instability of the aggregated cluster at large $D_s$, as observed in the H2 phase, is not predicted in the Keller-Segel model either. As the diffusion constant is larger, the gradient of the signal concentration pattern is smaller. If a cell can respond to any small signal gradient, as assumed in the continuum model, cells can aggregate even for any large $D_s$. On the other hand, in the present model of intracellular dynamics, there exists a minimum value of the gradient in signal concentration required for a cell to respond. Indeed, this value depends on the sharpness of the change of tumbling frequency against $c_u$, i.e., the value $\kappa_{\Delta} $ in eq.(\ref{eq_tumble}). As long as $\kappa_{\Delta}$ is finite, there exists minimum slope, which gives a maximum value of $D_s$ to make aggregation possible. Thus, the H2 phase exists as long as $\kappa_{\Delta} $ is finite. This value $\kappa_{\Delta}$ corresponds to the Hill coefficient in cell biology, and with its increase eq.(\ref{eq_tumble}) approaches a step function. As long as the Hill coefficient is finite, even if it is large, the H2 phase exists at large $D_s$, in contrast to the case with the Keller-Segel model.

\begin{figure}
\begin{center}
\scalebox{0.27}{\includegraphics{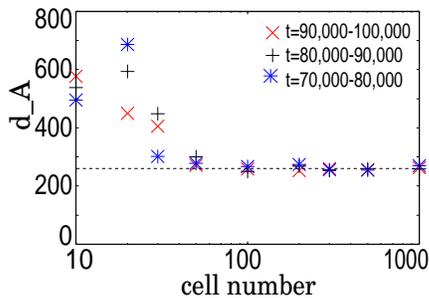}}
\caption{(Color online) Dependence of the cluster size (estimated by $d_A$) on the number
of cells. Parameters are $\tau _s=5.0 , \tau _a=50.0, \tau ^* =70.0 ,\sigma =0.0005, \nu =0.002, D_s=200. $ The cluster size is estimated by $d_A$, computed by the averages from 70,000 to 80,000 ($*$), from 80,000 to 90,000 ($+$), and from 90,000 to 100,000(x). Values from three temporal regions are computed to check the stability of the aggregated cluster.}
    \label{pic_size_cellnum}
\end{center}
\end{figure}

Finally, we study the dependence of the cluster size upon the number of cells at the aggregation phase. As plotted in Fig.\ref{pic_size_cellnum}, the cluster size (computed by $d_A$) is independent of the number of cells, as long as it is large enough to form a stable cluster (In the figure, the number is about 50).  

Even when the cell number is large, the cellular density in the cluster is sufficiently low 
(0.01 $ \mu m^2$ and a single bacterium is about $ 2\sim 3 \mu m $ in length\cite{Mittal_pnas03}), compared with the colony pattern, and so cells do not collide with each other and move independently. Owing to this independency, the cluster size is determined by the length beyond which the cell returns to the original cluster, given by $\lambda_n$, which is independent of the number of cells. In fact, Mittal et al. reported that the size of formed bacterial cluster is independent of the number of bacteria contained in it\cite{Mittal_pnas03}. Our numerical result agrees with their experiment.

In the present Letter, we have obtained conditions for chemotactic aggregation. One is the condition for decay of attractant in the medium, given by $\nu < \nu_c$, which is also derived from the continuum limit model, the so-called Keller-Segel model. The other concerns the inequality between the diffusion scale of the signal molecule within its lifetime and the motility scale of the random walk of cells, $\lambda _s > \lambda _n $. This latter condition, in addition to the condition for diffusion constant of signal molecule to be detected by the signal transduction, is not obtained in the continuum limit model. These conditions, as well as the Oosawa condition for chemotaxis, are general, and can be tested experimentally by varying the nature of the medium and signal molecules and by adopting mutants. As the cluster size constancy against cell density agrees with experimental data, experimental verifications of the predicted phases will be promising. In particular, partial aggregation may underlie intermittent expansion of cellular aggregates\cite{Shapiro_Oxfbook97}.

In the present model, the secretion of attractant from cells is independent of the intracellular state. It will be an important future issue to consider state-dependent secretion of chemicals and/or richer intracellular dynamics, to find complex spatiotemporal patterns\cite{BudreneBerg_nature91} as well as differentiation of intracellular states\cite{Shapiro_Oxfbook97}.

The authors would like to thank S.Sawai, S.Ishihara, K.Fujimoto, V.Nanjundiah, T.Shibata, for their valuable comments.
This work is supported by the JSPS Research Fellowships for Young Scientists.


\begin{thebibliography}{99}
\bibitem{com1}
The concentration here is defined as the deviation from the steady-state value so that it can be negative.
\bibitem{remark}
By scaling properly, we can fix the value of $\sigma$ by transforming 
the parameters $\nu \leftarrow \nu /\sigma$ and $D_s \leftarrow D_s /\sigma$.
\bibitem{eq_poisson}
Assuming that the tumbling occurs as a Poisson process, we use the coefficient of $\sqrt{2}$.
\end{thebibliography}
\end{document}